\documentclass[final,5p,times,twocolumn,sort&compress]{elsarticle}

\pdfoutput=1
\usepackage{amsmath,amssymb}
\usepackage[utf8]{inputenc}
\usepackage[T1]{fontenc}
\usepackage{hyperref}
\usepackage{multirow}
\usepackage{ifthen}

\newcommand{\Dmq}{\Delta m^2}
\newcommand{\eVq}{\ensuremath{\text{eV}^2}}
\newcommand{\Nuc}[2][]{{\ensuremath{\ifthenelse{\equal{#1}{}}{}{\mbox{}^{#1}}\text{#2}}}}

\makeatletter
\DeclareRobustCommand\recite[1]{\begingroup\@fileswfalse\cite{#1}\endgroup}
\makeatother

\graphicspath{{Figures/}}

\journal{Physics Letters B}

\begin{document}

\begin{frontmatter}
  
  \title{Solar Model Independent Constraints on the Sterile Neutrino
    Interpretation of the Gallium Anomaly}

  \author[af1,af2,af3]{M.~C.~Gonzalez-Garcia}
  \ead{concha.gonzalez-garcia@stonybrook.edu}
  \affiliation[af1]{\oraganization={Departament de F\'isica Qu\`antica i
      Astrof\'isica and Institut de Ci\`encies del Cosmos, Universitat
      de Barcelona,}, \addressline={Diagonal 647,}, \postcode={E-08028},
    \city={Barcelona,}, \contry={Spain}}
  \affiliation[af2]{\oraganization={Instituci\'o Catalana de Recerca i
      Estudis Avan\c{c}ats (ICREA)}, \addressline={Pg.\ Lluis Companys
      23}, \postcode={E-08010}, \city={Barcelona}, \country={Spain}}
  \affiliation[af3]{\oraganization={C.N.~Yang Institute for Theoretical
      Physics, Stony Brook University,}, \city={Stony Brook,},
    \state={NY}, \postcode={11794-3840,}, \country={USA}}

  \author[af4]{Michele Maltoni}
  \ead{michele.maltoni@csic.es}
  \affiliation[af4]{\organization={Instituto de F\'isica Te\'orica
      (IFT-CFTMAT), CSIC-UAM,}, \addressline={Calle de Nicol\'as Cabrera
      13--15, Campus de Cantoblanco,}, \postcode={E-28049,},
    \city={Madrid,}, \country={Spain}}

  \author[af1]{Jo\~ao Paulo Pinheiro}
  \ead{joaopaulo.pinheiro@fqa.ub.edu}

  \begin{abstract}
    We perform a global analysis of most up-to-date solar neutrino data
    and KamLAND reactor antineutrino data in the framework of the 3+1
    sterile neutrino mixing scenario (invoked to explain the results of
    the Gallium source experiments) with the aim of quantifying the
    dependence of the (in)compatibility of the required mixing with
    assumptions on the initial fluxes.  The analysis of solar data is
    performed in two alternative ways: using the flux predicted by the
    latest standard solar models, and in a model independent approach
    where the solar fluxes are also determined by the fit.  The
    dependence on the normalization of the capture rate in the solar
    Gallium experiments is also quantified.  Similarly, in the KamLAND
    analysis we consider both the case where the reactor flux
    normalization is assumed to be known \textit{a priori}, as well as a
    normalization free case which relies solely on available neutrino
    data.  Using a parameter goodness of fit test, we find that in most
    cases the compatibility between Gallium and solar+KamLAND data only
    occur at the $3\sigma$ level or higher.  We also discuss the
    implications of enforcing better compatibility by tweaking the
    mechanism for the energy production in the Sun.
  \end{abstract}

\end{frontmatter}

\section{Introduction}

It is almost two decades since the so-called \emph{Gallium
Anomaly}~\cite{Giunti:2006bj, Laveder:2007zz} became a standing puzzle
in neutrino physics.  In general terms, the anomaly accounts for the
deficit of the event rate measured in Gallium source experiments with
respect to the expectation.  It was originally observed in the
calibration of the gallium solar-neutrino detectors
GALLEX~\cite{GALLEX:1997lja, Kaether:2010ag} and
SAGE~\cite{SAGE:1998fvr, Abdurashitov:2005tb} with radioactive
\Nuc[51]{Cr} and \Nuc[37]{Ar} sources:
\begin{equation}
  \label{eq:detproc}
  \nu_e + \Nuc[71]{Ga} \to \Nuc[71]{Ge} + e^- .
\end{equation}
Using the detection cross section as predicted by
Bahcall~\cite{Bahcall:1997eg}, the average ratio of observed vs
predicted rates was found to be ${R}_{\text{GALLEX+SAGE}} = 0.88 \pm
0.05$~\cite{Abdurashitov:2005tb}, which represented a $2.4\sigma$
statistically significant deficit.
Most interestingly the Gallium Anomaly has been recently rechecked by
the BEST experiment~\cite{Barinov:2021asz, Barinov:2022wfh}, which
placed the \Nuc[51]{Cr} radioactive source at the center of a
concentric two-zone gallium target (thus effectively probing two
distinctive neutrino path lengths, of about 0.5~m and 1.1~m).  In both
zones they observe consistent deficits of $R_{\text{in}}= 0.79
\pm0.05$ and $R_{\text{out}}= 0.77 \pm0.05$~\cite{Barinov:2021asz,
  Barinov:2022wfh}, so the current combined level of the deficit is
$R_{\text{GALLEX+SAGE+BEST}} = 0.80 \pm0.05$~\cite{Barinov:2021asz,
  Barinov:2022wfh} promoting the statistical significance of the
anomaly beyond $4\sigma$.

Careful scrutiny~\cite{Elliott:2023xkb, Giunti:2022xat} of the
neutrino capture cross sections and its uncertainties does not provide
an explanation of the deficit, leaving open a possible effect in the
neutrino propagation.  The idea that $\nu_e$ may disappear during
propagation from source to detector is no surprise nowadays, as the
phenomenon of mass-induced neutrino flavour oscillations has been
established beyond doubt (see for example the review in
Ref.~\cite{ParticleDataGroup:2024cfk}) and the involved masses and
mixing are being determined with increasing accuracy by the combined
results of solar, reactor, atmospheric and long-baseline neutrino
experiments (see Ref.~\cite{Esteban:2024eli} for the latest global
analysis).  Unfortunately, it is precisely such accuracy which puts in
jeopardy the possible interpretation of the Gallium anomaly in terms
of neutrino oscillations.  Given the characteristic baseline
$\mathcal{O}(\text{meter})$ of the GALLEX, SAGE, and BEST radioactive
source experiments and their average neutrino energy
$\mathcal{O}(\text{MeV})$, a $\Dmq\gtrsim \mathcal{O}(\eVq)$ is
required to produce visible effects, and this is more than two orders
of magnitude larger than what indicated by the global analysis.  Hence
at least a fourth massive state must be involved in the propagation of
the neutrino ensemble with mass $m_4\sim\mathcal{O}(\text{eV})$.  This
in turn requires the introduction of a fourth neutrino weak
eigenstate, which must be an $SU(2)$ singlet to comply with the bounds
from the $Z$ invisible width~\cite{ParticleDataGroup:2024cfk}.  This
is how the light sterile neutrino scenario makes its entrance, but in
order to explain the Gallium anomaly in such way, the fourth state
must significantly mix with the three standard neutrinos,
$\sin^2\theta\sim \mathcal{O}(10)\%$.

The problem is that such large mixing would impact heavily the
oscillation signals included in the global analysis.  This results in
a strong tension between the sterile-neutrino interpretation of the
Gallium anomaly and other neutrino data~\cite{Berryman:2021yan,
  Goldhagen:2021kxe, Giunti:2022btk}.  In particular solar neutrinos
and reactor antineutrinos provide a clean test of the possible
projection of $\nu_e$ and $\bar \nu_e$ on a $\mathcal{O}(\text{eV})$
massive state.  Given the long baselines and the energies involved,
the oscillations driven by $\Dmq \sim \mathcal{O}(\eVq)$ are averaged
in both solar and KamLAND experiments so they directly test the mixing
relevant for the interpretation of the Gallium anomaly.  The analyses
presented in Refs.~\cite{Giunti:2022btk, Goldhagen:2021kxe} lead to
$2\sigma$ bounds $\sin^2\theta\lesssim 0.025$--$0.045$ of the
corresponding mixing angle, clearly disfavouring the sterile
oscillation interpretation of the Gallium anomaly.  Alternative
non-conventional scenarios have also been considered (see for example
Refs.~\cite{Brdar:2023cms, Farzan:2023fqa, Arguelles:2022bvt,
  Hardin:2022muu, Banks:2023qgd}), but they are nevertheless not free
from severe tension with other data~\cite{Giunti:2023kyo}.

It is worth noticing that the analyses of solar data in
Refs.~\cite{Giunti:2022btk, Goldhagen:2021kxe} were performed under
the assumption that the solar neutrino fluxes match the predictions of
the B16 Standard Solar Models (SSM)~\cite{Vinyoles:2016djt}.  As it is
known, over the last two decades the construction of Standard Solar
Models has suffered of the so called solar composition problem,
associated with the choice of the input values of heavy element
abundances embedded into the model.  Two approaches are usually
considered: either to rely on the older (and nowadays outdated)
results in Ref.~\cite{Grevesse1998} (GS98), which imply higher
metallicity and predict solar properties in good agreement with
helioseismology observations, or to use the newer abundances with
lower metallicity (obtained with more modern methodology and
techniques) summarized in Ref.~\cite{Asplund2009} (AGSS09) but which
do not agree with helioseismology.  This raises the question of the
possible solar model dependence of the derived bounds on the sterile
mixing.  Addressing this issue is the subject of this work.

In this respect, recently there have been two notable developments.
On the SSM building side, a new set of results
(MB22~\cite{Magg:2022rxb}, based on similar methodologies and
techniques as AGSS09 but with different atomic input data for the
critical oxygen lines, among other differences) led to a substantial
change in solar element abundances with respect to AGSS09, now more in
agreement with those from GS98.  Therefore, the models built following
MB22 provide a good description of helioseismology results.
Furthermore, as shown in Ref.~\cite{Gonzalez-Garcia:2023kva} it is
possible perform a solar model independent analysis of the latest
solar neutrino data (in combination with KamLAND) which allows for a
simultaneous determination of the oscillation parameters and the
normalization of the different components of the solar neutrino
fluxes.  The analysis in Ref.~\cite{Gonzalez-Garcia:2023kva} was
performed in the framework of $3\nu$ oscillations, but the same
methodology can be applied in the presence of mixing with a fourth
sterile neutrino state, thus determining the constraints on the
sterile interpretation of the Gallium anomaly in a way which is
completely independent of the modelling of the Sun.

This motivates the study which we present here with the following
outline.  In Sec.~\ref{sec:frame} we briefly summarize the different
elements entering in the analysis of solar and KamLAND data.
Section~\ref{sec:SSM} contains the results obtained in the framework
of the new generation of SSMs.  In Sec.~\ref{sec:free} we present the
results of our SSM independent analysis, subject solely to the
imposition of the luminosity constraint which links together the
neutrino flux and the thermal energy produced by each nuclear reaction
in the Sun (and accounts for the fact that the overall amount of
generated thermal energy must match the observed solar radiated
luminosity).  We also comment on the implications that assuming at
face value the sterile solution of the Gallium anomaly would have on
the mechanism for energy production in the Sun.  Finally in
Sec.~\ref{sec:conclu} we summarize our conclusions.

\section{Framework}
\label{sec:frame}

In the analysis of solar neutrino experiments we include the total
rates from the radiochemical experiments
Chlorine~\cite{Cleveland:1998nv} (1 data point),
Gallex/GNO~\cite{Kaether:2010ag} (2 data points), and
SAGE~\cite{Abdurashitov:2009tn} (1 data point), the spectral and
day-night data from phases I-IV of
Super-Kamiokande~\cite{Hosaka:2005um, Cravens:2008aa,
  Abe:2010hy,Super-Kamiokande:2023jbt} (44, 33, 42, and 46 data
points, respectively), the results of the three phases of SNO in the
form of the day-night spectrum data of SNO-I~\cite{Aharmim:2007nv} and
SNO-II~\cite{Aharmim:2005gt} and the three total rates of
SNO-III~\cite{Aharmim:2008kc} (34, 38, and 3 data points,
respectively),\footnote{This corresponds to the analysis labelled
\textsc{SNO-data} in Ref.~\cite{Gonzalez-Garcia:2013usa}, and is at
difference with Ref.~\cite{Gonzalez-Garcia:2023kva} which used instead
an alternative set (labelled \textsc{SNO-poly} in
Ref.~\cite{Gonzalez-Garcia:2013usa}) based on an effective
\emph{MSW-like} polynomial parametrization for the day and night
survival probabilities of the combined SNO phases I--III, as detailed
in Ref.~\cite{Aharmim:2011vm}.  The \textsc{SNO-poly} approach can be
efficiently applied to $3\nu$ oscillations but it relies on the
unitarity relation $P_{ee} + P_{e\mu} + P_{e\tau} = 1$ which does not
hold in the presence of extra sterile states.  This leads to small
quantitative differences between the results in
Ref.~\cite{Gonzalez-Garcia:2023kva} and those of this work in the
limit of very small sterile neutrino mixing.} and the full spectra
from Borexino Phase-I~\cite{Bellini:2011rx} (33 data points),
Phase-II~\cite{Borexino:2017rsf} (192 data points), and
Phase-III~\cite{BOREXINO:2022abl} (120 data points), together with
their latest results based on the Correlated Integrated Directionality
method~\cite{Borexino:2023puw} (1 data point; see
Refs.~\cite{Gonzalez-Garcia:2023kva, Coloma:2022umy} for details of
our analysis of Borexino II and III phases).
As mentioned in the introduction, an attempt to alleviate the Gallium
anomaly invokes the uncertainties of the capture cross
section~\cite{Berryman:2021yan, Giunti:2022xat, Elliott:2023xkb}.
This raises the issue of whether the estimated rate of solar neutrino
events in Gallium experiments is really robust.  To quantify the
impact of such possibility, we performed two variants of our solar
neutrino data analysis.  In the first one we include the results of
the solar Gallium experiments at their nominal values as provided by
the collaboration.  In the second one we introduce an additional
parameter, $f_\text{Ga}$, which is varied freely in the fits and
accounts for an overall scaling of the predicted event rates in the
solar Gallium experiments.  Statistically this is equivalent to
removing the Gallium experiments from the solar analysis, but done in
this way it allows us to quantify the range of $f_\text{Ga}$ favoured
by the solar data.  We will return to this point in the next section.

Concerning KamLAND, we include in the analysis the separate DS1, DS2,
DS3 spectral data~\cite{Gando:2013nba} (69 data points).  Since
KamLAND has no near detector, the theoretical uncertainties in the
calculations of the reactor neutrino spectra should be carefully taken
into account.  Here we will consider two limiting cases.  In the first
one we will assume some theoretically calculated reactor fluxes which
will be used as input in the analysis of the KamLAND data.  We will
label this analysis as <<reactor flux constrained>>, or
<<KamLAND-RFC>> in short.  For concreteness we use as
\emph{theoretical reactor fluxes} those predicted by an ad-hoc model
adjusted to perfectly reproduce the spectrum observed in Daya-Bay
experiment~\cite{DayaBay:2021dqj} (and their uncertainties) in the
absence of sterile oscillations.  Similar results would be obtained
with any of the reactor flux models in the literature as long as they
are consistent with the Daya-Bay measurements in the framework of
$3\nu$ mixing.  By construction this is the most limiting scenario.
Alternatively one can use the Daya-Bay reactor spectra as a truly
experimental input, taking into account that at the moment of their
detection the suppression induced by the $\Dmq\sim \mathcal{O}(\eVq)$
oscillations has already taken place (so that the neutrino flux
generated by the reactor cores must be proportionally larger).  This
is a more conservative scenario in which nothing is assumed about the
theoretical prediction of the reactor flux normalization.  We label
this analysis as <<reactor flux free>>, or <<KamLAND-RFF>> in short.

In what respects the relevant survival probabilities, we focus here on
a 3+1 scenario where $\{\nu_1, \nu_2, \nu_3\}$ (with mass-squared
splittings $\Dmq_{21}$ and $\Dmq_{31}$ as determined by the standard
$3\nu$ oscillation analysis) are dominantly admixtures of the
left--handed states $\{\nu_e, \nu_\mu, \nu_\tau \}$, and a fourth
massive state $\nu_4$ (with a mass-squared splitting $\Dmq_{41} \simeq
\Dmq_{42} \simeq \Dmq_{43} \sim \mathcal{O}(\eVq)$) is mostly sterile
(\textit{i.e.}, not coupled to the weak currents) but has some
non-vanishing projection over the left-handed states (see appendix C
of Ref.~\cite{Kopp:2013vaa} for details).  We obtain the oscillation
probabilities for solar neutrinos by numerically solving the evolution
equation for the neutrino ensemble from the neutrino production point
to the detector including matter effects both in the Sun and in the
Earth, with no other approximation than the assumption that the
evolution in the Sun is adiabatic.
We parametrize the mixing matrix as in Ref.~\cite{Kopp:2013vaa,
  Dentler:2018sju}:
\begin{equation}
  U = V_{34} V_{24} V_{14} V_{23} V_{13} V_{12} \,,
\end{equation}
where $V_{ij}$ is a rotation in the $ij$ plane by an angle
$\theta_{ij}$, which in general can also contain a complex phase (see
appendix~A of~\cite{Kopp:2013vaa} for a discussion).  Following
Ref.~\cite{Goldhagen:2021kxe} we make use of the fact that that bounds
from $\nu_\mu$ disappearance in atmospheric and long-baseline
neutral-current measurements render the solar neutrino data
effectively insensitive to $\theta_{34}$ and $\theta_{24}$, hence in
what follows we set $\theta_{34} = \theta_{24} = 0$ and obtain:
\begin{equation}
  \label{eq:U}
  U =
  \begin{pmatrix}
    c_{14} c_{13} c_{12} & c_{14} c_{13} s_{12} & c_{14} s_{13} & s_{14} \\
    \cdot & \cdot & \cdot & 0 \\
    \cdot & \cdot & \cdot & 0 \\
    -s_{14} c_{13} c_{12} & -s_{14} c_{13} s_{12} & -s_{14} s_{13} & c_{14}
  \end{pmatrix}.
\end{equation}
Under these approximations and taking into account that for the
distance and energies of solar and KamLAND neutrinos the oscillations
driven by $\Dmq_{31}$ and $\Dmq_{4i}$ are averaged out, the relevant
probabilities depend only on the three angles $\theta_{12}$,
$\theta_{13}$, $\theta_{14}$ as well as $\Dmq_{21}$.  Furthermore
in~\cite{Kopp:2013vaa} it has been shown that the determination of
$\theta_{13}$ is basically unaffected by the presence of a sterile
neutrino, so we can fix it to the best fit value $s_{13}^2 = 0.02224$
obtained in the $3\nu$ scenario and safely neglect its current
uncertainty.  Altogether the relevant probabilities depend on three
parameters: $\Dmq_{21}$, $\sin^2\theta_{12}$, and $\sin^2\theta_{14}$.

Finally, in the analysis of the Gallium source experiments one can set
$\Dmq_{21} = \Dmq_{31} = 0$, so that the corresponding $\nu_e$
survival probability reduces to the well-known $2\nu$ vacuum
oscillation formula and involves only two parameters, namely the large
mass-squared splitting $\Dmq_{41} = \Dmq_{42} = \Dmq_{43}$ and the
mixing angle $\theta_{14}$:
\begin{equation}
  P_{ee}^\text{Ga-source} =
  1 - \sin^2(2\theta_{14}) \sin^2\bigg( \frac{\Dmq_{41} L}{4E} \bigg) \,.
\end{equation}
Hence the only common parameter between the Gallium-source experiments
and the solar and KamLAND experiments is $\theta_{14}$.  In what
follows we will perform several compatibility tests of the oscillation
parameters allowed by solar data (both alone and in combination with
KamLAND) and those implied by the analysis of the Gallium source
experiments.  To this aim we will make use of a
$\Delta\chi^2_\text{Ga-source}(\theta_{14})$ function inferred from
the combined fit of the Gallium source experiments presented in
Ref.~\cite{Barinov:2022wfh}.

\section{Results}
\label{sec:res}

\begin{figure*}[t]\centering
  \includegraphics[width=0.94\textwidth]{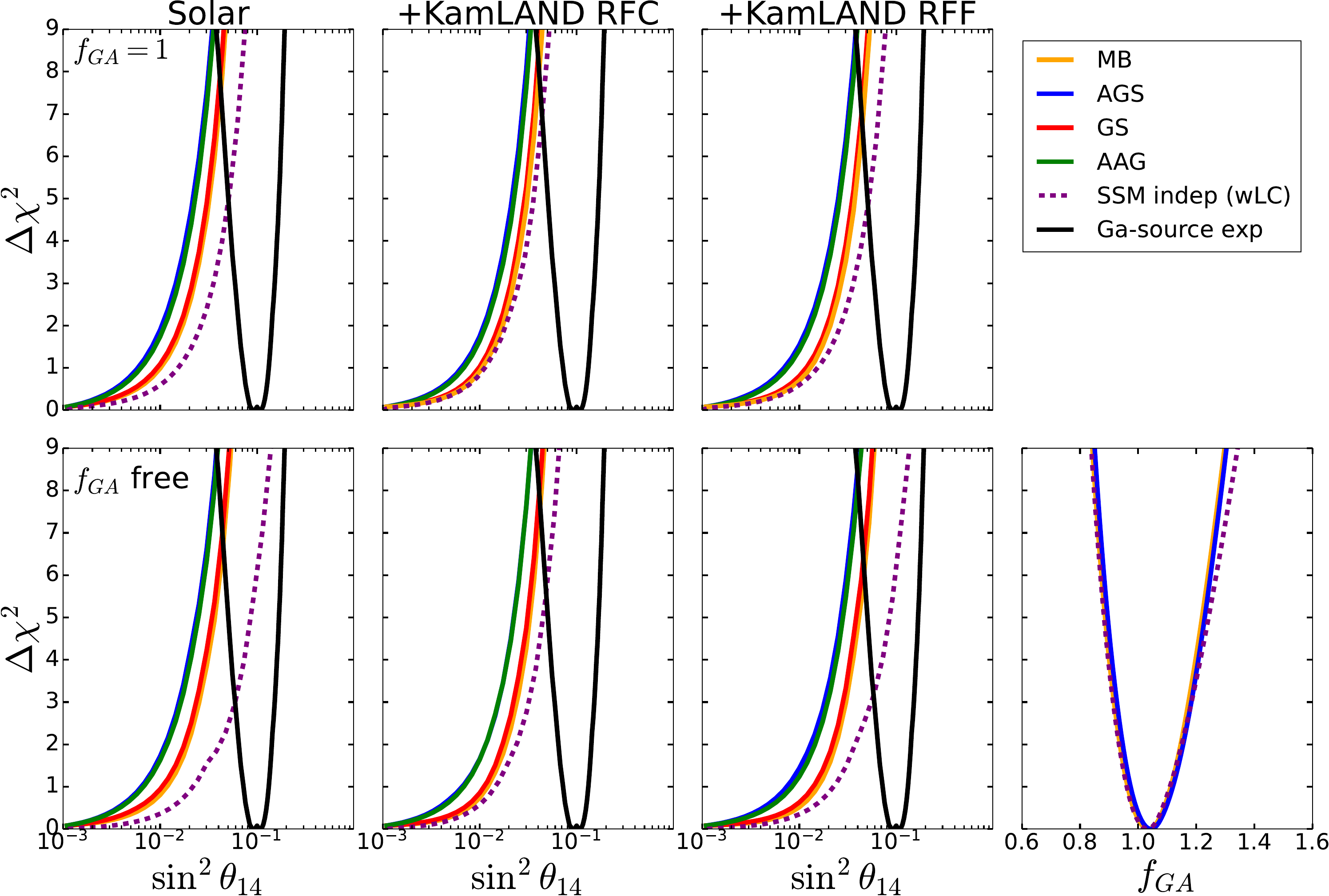}
  \caption{One-dimensional projection of the global $\Delta\chi^2$ on
    the mixing angle $\theta_{14}$ after marginalization over the
    undisplayed parameters, for different assumptions on the solar
    neutrino fluxes as labelled in the legend.  The first (second)
    [third] column corresponds to the analysis including solar-only
    (solar+KamLAND-RFC) [solar+KamLAND-RFF] data.  The upper (lower)
    panels show the results with $f_\text{Ga}=1$ (free $f_\text{Ga}$).
    In all panels the black parabola is
    $\Delta\chi^2_\text{Ga-source}(\theta_{14})$ as inferred from the
    combined analysis of the Gallium source experiments presented in
    Ref.~\cite{Barinov:2022wfh}.  In the rightmost lower panel we show
    the dependence of $\Delta\chi^2$ on the normalization parameter
    $f_\text{Ga}$.}
  \label{fig:ssmchi}
\end{figure*}

\subsection{Updated bounds with SSM fluxes}
\label{sec:SSM}

In the first round of analyses we present the results of our fits to
solar data in the framework of four different versions of the B23
standard solar models.  Concretely, we consider the SSMs computed with
the abundances compiled in table 5 of~\cite{Magg:2022rxb} based on the
photospheric solar mixtures (MB-phot; the results obtained with
meteoritic solar mixtures are totally equivalent), as well as models
with the solar composition taken from Ref.~\cite{Asplund2021} (AAG21),
from the meteoritic scale of Ref.~\cite{Asplund2009} (AGSS09-met), and
from Ref.~\cite{Grevesse1998} (GS98).

The results of the SSM constrained analysis are presented in
Fig.~\ref{fig:ssmchi} where we plot $\Delta\chi^2(\theta_{14})$ for
different choices of the solar fluxes and of the KamLAND analysis.
For sake of comparison we also show
$\Delta\chi^2_\text{Ga-source}(\theta_{14})$ as inferred from the
combined fit of the Gallium source experiments presented in
Ref.~\cite{Barinov:2022wfh}.
As can be seen all variants of the solar (+KamLAND) data analysis
favour $\theta_{14}=0$, so that the fit always results into an upper
bound on the allowed range of $\sin^2\theta_{14}$ which is in clear
tension with the values required to explain the Gallium anomaly.
Comparing the upper and lower panels we also see that relaxing the
constraint on the normalization parameter $f_\text{Ga}$ does not lead
to any significant difference in the outcome.  Let us notice that
$f_\text{Ga}$ is a factor introduced only in the fit of solar
neutrinos, without affecting the analysis of the Gallium source
experiments, because in here we are interested in testing if relaxing
some assumptions in the solar and KamLAND analysis can lead to a
better agreement with the sterile neutrino interpretation of the
results of Gallium source experiments.  Studies of correlated effects
affecting the Gallium capture rate in both solar and source
experiments, such as varitions of the capture cross sections, have
been presented in various works (see, \textit{e.g.},
Refs.~\cite{Bahcall:1997eg, Haxton:1998uc,
  Frekers:2015wga,Kostensalo:2019vmv, Semenov:2020xea,
  Giunti:2022xat,Elliott:2023xkb}).  Statistically, allowing
$f_\text{Ga}$ to be free in the solar and KamLAND analysis is
equivalent to removing the solar Gallium experiments from the fit, as
clearly visible in Fig.~3 of Ref.~\cite{Gonzalez-Garcia:2023kva} where
both approches have been explicitly implemented.  This is due to the
lack of spectral and day-night capabilities in Gallium data, which
prevents them from providing further information beyond the overall
normalization scale of the signal.  The advantage of performing the
analysis introducing the unconstrainted $f_\text{Ga}$ factor, is that
one can quantify the range of $f_\text{Ga}$ favoured by the fit.
Interestingly from the bottom-right panel we see that the solar and
KamLAND data always favours $f_\text{Ga}$ close to one with $1\sigma$
uncertainty of about $7\%$.  In comparison, the prediction of the
neutrino capture cross section in Gallium in the different models can
vary up to about $15$\%~\cite{Bahcall:1997eg, Haxton:1998uc,
  Frekers:2015wga, Kostensalo:2019vmv, Semenov:2020xea,
  Giunti:2022xat, Elliott:2023xkb}.  This means that the global
analyses of solar experiments do not support a significant
modification of the neutrino capture cross section in Gallium, or any
other effect inducing an energy independent reduction of the detection
efficiency in the solar experiments.

\begin{table*}[t]\centering
  \catcode`?=\active\def?{\hphantom{0}}
  \renewcommand{\arraystretch}{1.2}
  \begin{tabular}{|l|l|ccc|ccc|}
    \hline
    \multicolumn{2}{|c|}{}
    & \multicolumn{3}{c|}{$f_\text{Ga}=1$}
    & \multicolumn{3}{c|}{$f_\text{Ga}$ free}
    \\
    \hline
    & \hfil SSM
    & $\chi^2_\text{PG} / n$ & $p$-value ($\times 10^{-3}$) & $\#\sigma$
    & $\chi^2_\text{PG} / n$ & $p$-value ($\times 10^{-3}$) & $\#\sigma$
    \\
    \hline
    \multirow{3}{*}{Solar}
    & MB-phot/GS98 & 14.9 & 0.11 & 3.9 & 13.1 & 0.3? & 3.6
    \\
    & AAG21/AGSS09 & 18.7 & 0.2? & 4.3 & 17.3 & 0.03 & 4.2
    \\
    & SSM indep (wLC) & ?9.1 & 2.6? & 3.0 & ?4.9 & 27?  & 2.2
    \\
    \hline
    \multirow{3}{*}{\begin{tabular}{@{}l@{}}Solar +\\[2pt]KL-RFC\end{tabular}}
    & MB-phot/GS98 & 15.9 & 0.07 & 4.0 & 15.1 & 0.1? & 3.9
    \\
    & AAG21/AGSS09 & 19.4 & 0.1? & 4.4 & 18.7 & 0.01 & 4.3
    \\
    & SSM indep (wLC) & 13.5 & 0.23 & 3.7 & 10.5 & 1.2? & 3.2
    \\
    \hline
    \multirow{3}{*}{\begin{tabular}{@{}l@{}}Solar +\\[2pt]KL-RFF\end{tabular}}
    & MB-phot/GS98 & 13.2 & 0.28 & 3.6 & 11.7 & 0.64 & 3.4
    \\
    & AAG21/AGSS09 & 17.3 & 0.03 & 4.2 & 16.0 & 0.06 & 4.0
    \\
    & SSM indep (wLC) & ?8.7 & 3.1? & 2.9 & ?4.8 & 29?  & 2.2
    \\
    \hline
  \end{tabular}
  \caption{Results of the PG test for the different solar flux model
    assumptions and the different analysis variants.}
  \label{tab:ssmpg}
\end{table*}

The quantitative question to address is the level of (in)compatibility
of these results from the solar (+KamLAND) analysis with those from
the Gallium source experiments in the context of the 3+1 scenario.
Consistency among different data sets can be quantified with the
parameter goodness-of-fit (PG) test~\cite{Maltoni:2003cu}.  For a
number $N$ of uncorrelated data sets $i$, each one depending on $n_i$
model parameters and collectively depending on $n_\text{glob}$
parameters, it can be shown that the test statistic
\begin{equation}
  \label{eq:PGtest}
  \chi^2_\text{PG}
  \equiv \chi^2_\text{min,glob} - \sum_i^N \chi^2_{\text{min}, i}
  = \min\bigg[ \sum_i^N\chi^2_i \bigg] - \sum_i \min\chi^2_i
\end{equation}
follows a $\chi^2$ distribution with $n_\text{PG} \equiv \sum_i n_i -
n_\text{glob}$ degrees of freedom~\cite{Maltoni:2003cu}.  In this
section we have $N=2$, and the relevant number of parameters are
$n_\text{solar(+KamLAND)} = 3$ (or $4$ for analysis with $f_\text{Ga}$
free), $n_\text{Ga-source} = 2$ and $n_\text{global} = 4$ (or $5$ for
analysis with $f_\text{Ga}$ free).  So for all tests $n_\text{PG} =
1$, reflecting the fact that the only parameter in common between the
solar (+KamLAND) and the Gallium-source data sets is $\theta_{14}$.
We list in table~\ref{tab:ssmpg} the results of applying the PG test
to the different variants of the analysis.  As can be seen, from the
results in the lines labeled MB-phot/GS98 (AAG21/AGSS09) in the
``Solar'' case, the analysis of the solar neutrino data performed in
the framework of the SSM's with higher (lower) metallicity are
incompatible with the Gallium source experiments at the $3.6\sigma$
($4.2\sigma$) level even when allowing for a free $f_\text{Ga}$.
Looking at the corresponding lines for the ``Solar+KL-RFC'' and
``Solar+KL-RFF'' fits we see that combination with KamLAND data
results into a slight improvement or weakening of the incompatibility
depending on the assumption on the reactor fluxes.

To further illustrate the interplay between solar and KamLAND data we
show in Fig.~\ref{fig:triangle} the two-dimensional projection of the
global $\Delta\chi^2$ for the separate analysis of solar-only and
KamLAND-only results under different assumptions for the corresponding
input fluxes.  From the upper left panel, as expected, we see that the
determination of $\Dmq_{21}$ in KamLAND is robust irrespective of the
presence of sterile neutrinos or the assumptions on the reactor flux
normalization, since both occurrences only affect the overall scale of
the signal whereas $\Dmq_{21}$ is determined by the distortions of the
energy spectral shape.  As it is well known from the results of the
$3\nu$ analysis, the $\Dmq_{21}$ values favoured by KamLAND lie in the
upper $2\sigma$ allowed range of the solar neutrino fit.  From the
upper right panel we see that the dependence of
$\Delta\chi^2_\text{solar}$ on $\theta_{14}$ within the $\Dmq_{21}$
interval favoured by KamLAND is flatter than at the lower $\Dmq_{21}$
values preferred by the solar-only analysis, which means that the
solar-only bound on $\theta_{14}$ becomes weaker when $\Dmq_{21}$ is
constrained to the KamLAND range.  In addition, the lower panel shows
that in the <<KamLAND-RFF>> analysis (for which no information on the
absolute reactor flux normalization is included) there is a degeneracy
between $\theta_{12}$ and $\theta_{14}$.  Such degeneracy is expected
as the KamLAND survival probability in vacuum reads
\begin{multline}
  P_{ee}^\text{KamLAND} \simeq
  \cos^4\theta_{14} (\cos^4\theta_{13} + \sin^4\theta_{13})
  + \sin^4\theta_{14}
  \\
  - \cos^4\theta_{14}\, \cos^4\theta_{13} \,
  \sin^2(2\theta_{12})\, \sin^2(\Dmq_{21} L/E)
\end{multline}
so the spectral shape of the signal only provides information on the
ratio of the energy-dependent and energy-independent pieces
\begin{equation}
  \label{eq:ratio}
  \frac{\cos^4\theta_{14}\, \cos^4\theta_{13} \sin^2(2\theta_{12})}
       {\cos^4\theta_{14}(\cos^4\theta_{13} + \sin^4\theta_{13})
         + \sin^4\theta_{14}}
\end{equation}
whose isocontours precisely trace the magenta lines in the
($\sin^2\theta_{12}$, $\sin^2\theta_{14}$) plane observed in the lower
panel.  As a consequence of all this, the combination of
solar+KamLAND-RFF data leads to a slight weakening of the bounds on
$\theta_{14}$ compared to the solar-only analysis, as can be seen
comparing the left and right panels in Fig.~\ref{fig:ssmchi} as well
as the corresponding values of the PG test in Table~\ref{tab:ssmpg}.
On the contrary the analysis of KamLAND with constrained reactor
fluxes can independently bound both the numerator and denominator in
Eq.~\eqref{eq:ratio} and therefore provides an additional constraint
on $\theta_{14}$, as illustrated by the filled green regions in the
lower panel of Fig.~\ref{fig:triangle}.  In the end when combining
solar and KamLAND-RFC this second effect overcompensates the weakening
of the solar bound associated with the larger $\Dmq_{21}$ value, so
that the solar+KamLAND-RFC analysis results in a stronger
$\theta_{14}$ bound than the solar-only fit.

\subsection{Bounds from solar model independent analysis}
\label{sec:free}

Let us now discuss the results of fits performed without the
assumption of standard solar model fluxes (but still retaining the
condition of consistency with the observed solar luminosity).
As mentioned in the introduction, the combined analysis of present
solar neutrino experiments and KamLAND reactor data allows for the
simultaneous determination of the relevant oscillation parameters
together with the normalizations $\Phi_i$ of the eight solar neutrino
fluxes~--- five produced in the reactions of the pp-chain, $i \in
\big\lbrace \Nuc{pp},\, \Nuc[7]{Be},\, \Nuc{pep},\, \Nuc[8]{B},\,
\Nuc{hep} \rbrace$, and three originating from the CNO-cycle, $i \in
\big\lbrace \Nuc[13]{N},\, \Nuc[15]{O},\, \Nuc[17]{F} \big\rbrace$.
In Ref.~\cite{Gonzalez-Garcia:2023kva} we presented such determination
in the framework of $3\nu$ oscillations, to which we refer for the
technical details.  In brief, in this kind of analysis the flux
normalizations are allowed to vary freely subject only to a minimal
set of physical constraints and working assumptions, the most relevant
of which are:
\begin{itemize}
\item the fluxes must be positive: $\Phi_i \geq 0$;

\item the number of nuclear reactions terminating the pp-chain should
  not exceed the number of nuclear reactions which initiate
  it~\cite{Bahcall:1995rs, Bahcall:2001pf}: $\Phi_{\Nuc[7]{Be}} +
  \Phi_{\Nuc[8]{B}} \leq \Phi_{\Nuc{pp}}
  + \Phi_{\Nuc{pep}}$;

\item the ratio of the \Nuc{pep} neutrino flux to the \Nuc{pp}
  neutrino flux is fixed to high accuracy because they have the same
  nuclear matrix element, hence it is constrained to lie within a
  narrow range of those of the SSMs: $\Phi_{\Nuc{pep}} \,\big/\,
  \Phi_{\Nuc{pp}} = (1.004 \pm 0.018)\, \Phi^\text{GS98}_{\Nuc{pep}}
  \,\big/\, \Phi^\text{GS98}_{\Nuc{pp}}$;

\item in what respects the CNO fluxes, we have verified that for the
  assessment of the (in)compatibility with the Gallium source
  experiments the most relevant fluxes are those of the pp-chain, and
  the results are very insensitive to the assumptions on the CNO
  fluxes.  Thus, for simplicity, we have assumed a common rescaling of
  the three CNO fluxes with respect to those predicted by the SSM's.
\end{itemize}
In addition, the so-called ``luminosity constraint'' (\textit{i.e.},
the requirement that the overall amount of thermal energy generated
together with each neutrino flux matches the observed solar
luminosity~\cite{Spiro:1990vi}) implies that
\begin{equation}
  \label{eq:lumsum2}
  \frac{L_\odot}{4\pi \, (\text{A.U.})^2}
  = \sum_{i=1}^8 \alpha_i \Phi_i
  \equiv \frac{L_\odot(\text{$\nu$-inferred})}{4\pi \, (\text{A.U.})^2}
\end{equation}
where $\alpha_i$ is the energy released by the nuclear fusion
reactions associated with the $i^\text{th}$ neutrino flux; its
numerical value ranges from 13.099~MeV for $i = \Nuc{pp}$ down to
3.755~MeV for $i = \Nuc{hep}$, and is independent of the details of
the solar model to an accuracy of $10^{-4}$ or
better~\cite{Bahcall:2001pf, 2021JPhG...48a5201V}.  $\Phi_{\Nuc{pp}}$
is also the largest flux and by itself it contributes about $\sim
92\%$ of $L_\odot(\text{$\nu$-inferred})$.  In Eq.~\eqref{eq:lumsum2}
$L_\odot$ denotes the Sun luminosity as directly extracted from the
available satellite data: $L_\odot \,\big/\, [4\pi \, (\text{A.U.})^2]
= 8.4984 \times 10^{11} \, \text{MeV} \, \text{cm}^{-2} \,
\text{s}^{-1}$~\cite{ParticleDataGroup:2024cfk}, with an uncertainty
of $0.34\%$ due to systematics.  Technically, the luminosity
constraint is imposed by adding a prior to $\chi^2_\text{solar}$:
\begin{equation}
  \label{eq:priorLC}
  \chi^2_\text{LC} = \bigg(
  \frac{L_\odot(\text{$\nu$-inferred}) \,\big/\, L_\odot - 1}{0.0034}
  \bigg)^2 \,.
\end{equation}

\begin{figure}[t]\centering
  \includegraphics[width=\linewidth]{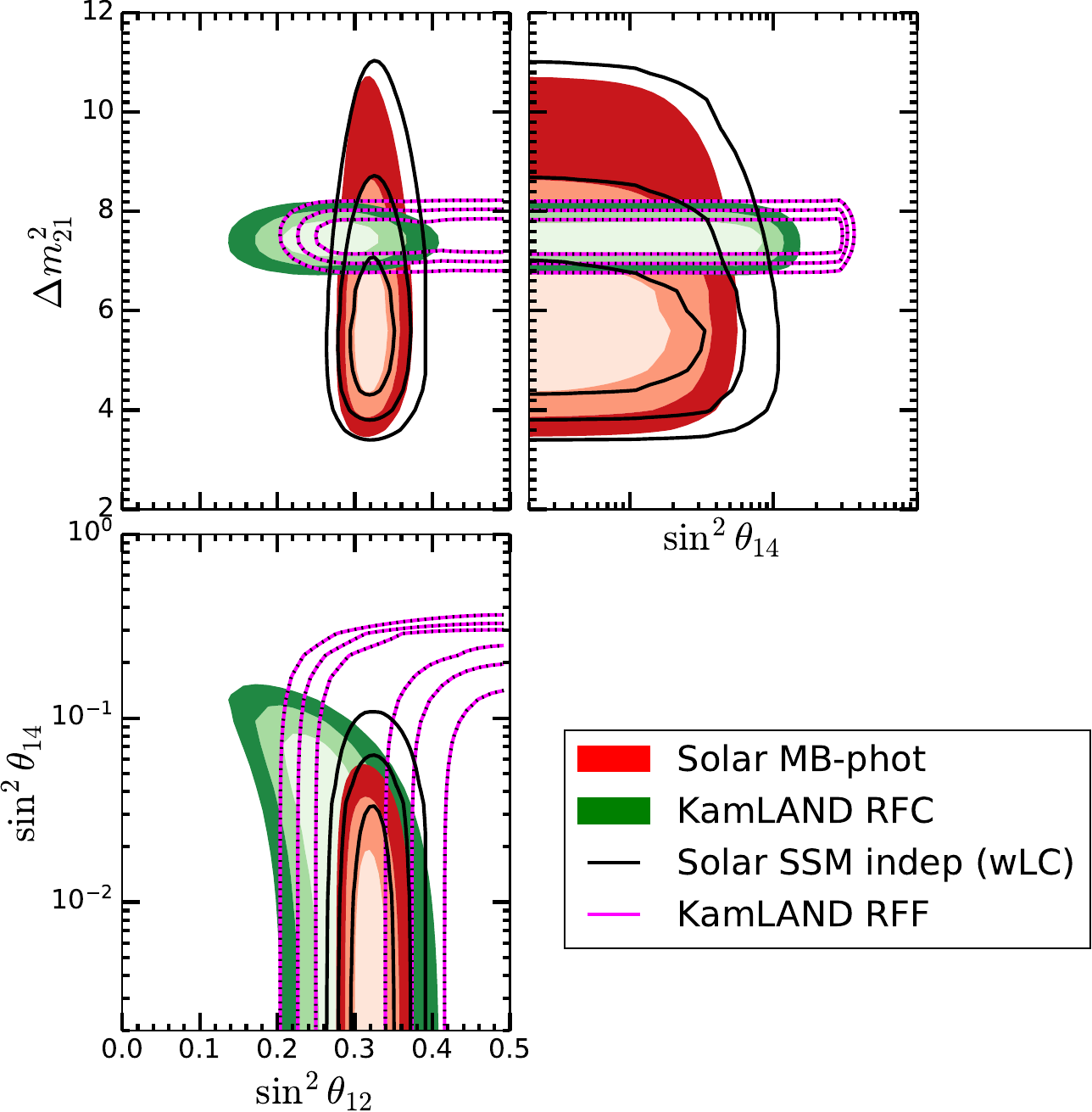}
  \caption{Two-dimensional projection of the global $\Delta\chi^2$
    (for $f_\text{GA}=1$) on
    the relevant oscillation parameters at $1\sigma$, $2\sigma$ and
    $3\sigma$ after marginalization over the undisplayed parameters.
    The full red regions (void black contours) correspond to the
    analysis of solar data with MB-phot (free) fluxes, while the full
    green regions (void magenta contours) correspond to the KamLAND-RFC
    (KamLAND-RFF) fit.}
  \label{fig:triangle}
\end{figure}

In Fig.~\ref{fig:triangle} we show as void black contours the
constraints on the oscillation parameters from our SSM independent
analysis of solar data (labeled as ``SSM indep (wLC)'' in short), as
obtained after marginalization over the various flux normalizations.
Comparing these contours with the full red regions (corresponding to
the analysis with solar fluxes as predicted by the MB-phot SSM) we see
that the determination of the oscillation parameters is only slightly
loosened in the SSM independent fit, and in particular the analysis
still yields a strong bound on $\theta_{14}$.
The corresponding one-dimensional projections
$\Delta\chi^2(\theta_{14})$ (both for solar-only and in combination
with the two variants of the KamLAND fit) are shown as dotted lines in
the various panels of Fig.~\ref{fig:ssmchi}, and the values of the PG
tests are given in Table~\ref{tab:ssmpg}.  From the table we read
that, as long as the normalization parameter $f_\text{Ga}$ is kept to
its nominal value $f_\text{Ga}=1$, the level of (in)compatibility
between the solar (+KamLAND) data and the Gallium source experiments
is at a level $\gtrsim 3\sigma$.  The tension can only be relaxed to
the $\sim 2.2\sigma$ level when allowing the value of $f_\text{Ga}$ to
float freely.

\subsection{Implications for the $\nu$-inferred Sun Luminosity}

\begin{figure*}[t]\centering
  \raisebox{1.2mm}{\includegraphics[width=0.48\textwidth]{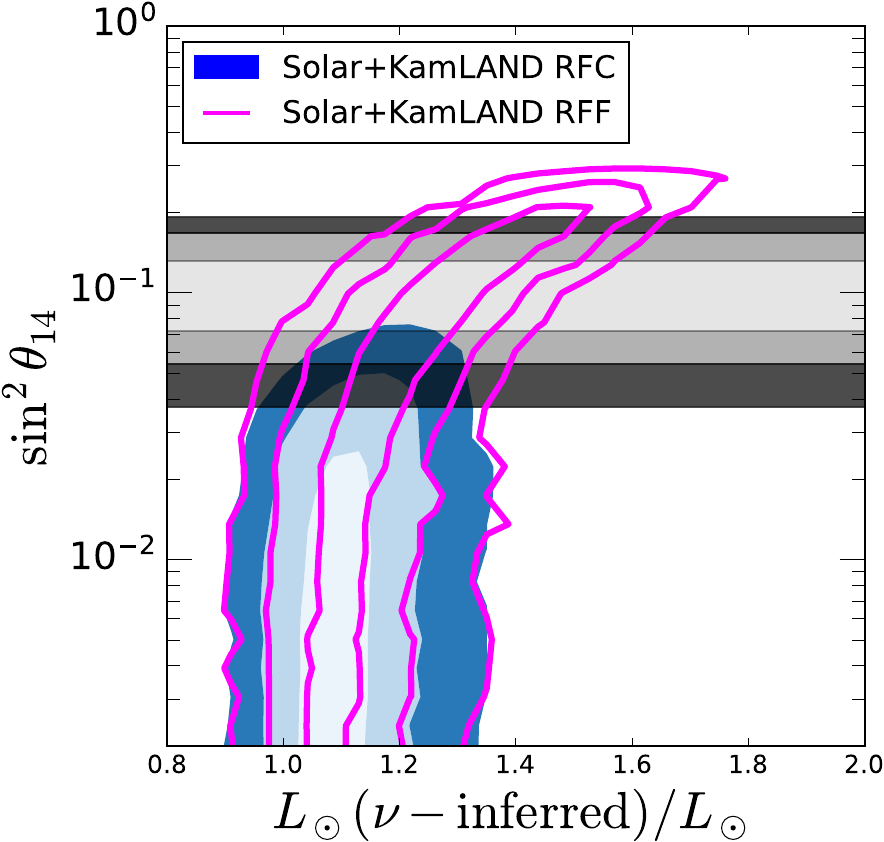}}
  \hfil\includegraphics[width=0.461\textwidth]{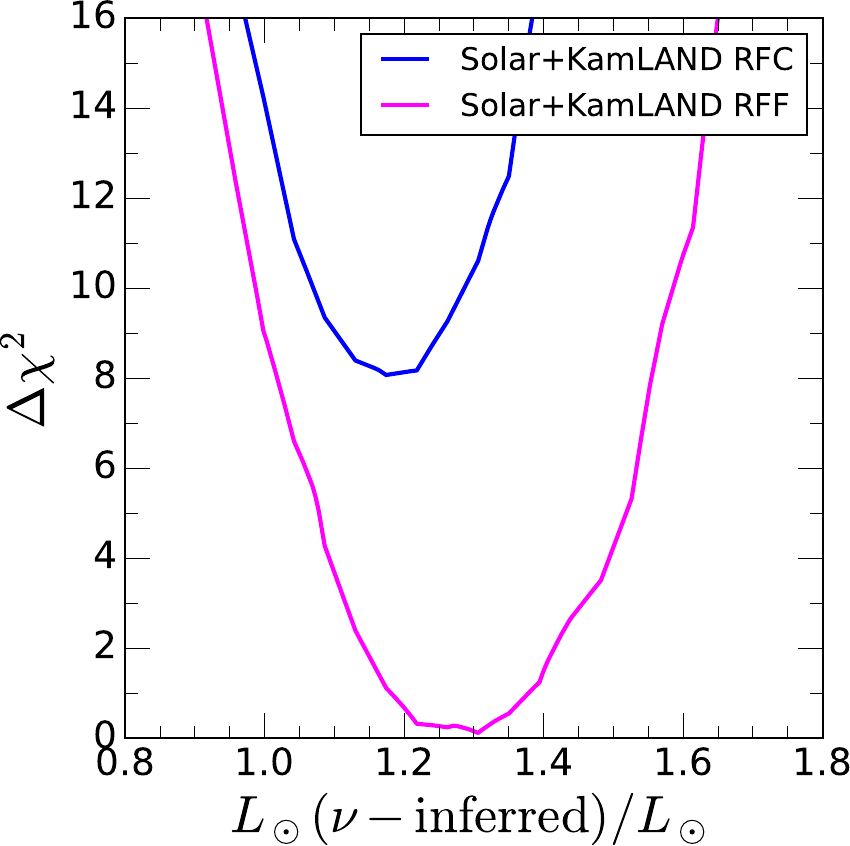}
  \caption{Left: dependence of the $1\sigma$, $2\sigma$, $3\sigma$
    ranges of $\sin^2\theta_{14}$ from the analysis of solar+KamLAND
    without imposing the constraint in Eq.~\eqref{eq:priorLC} on the
    resulting neutrino-inferred solar luminosity.  Fill (void) regions
    correspond to solar+KamLAND-RFC (solar+KamLAND-RFF) analysis.  The
    horizontal grey regions illustrate the $1\sigma$, $2\sigma$,
    $3\sigma$ ranges required to explain the Gallium source results.
    Right: value of $\Delta\chi^2$ from the joint analysis of
    solar+KamLAND and Ga-source data, defined with respect to the sum
    of the two separate best-fit $\chi^2_\text{min}$, as a function of
    the neutrino-inferred solar luminosity.  In these analysis we keep
    $f_\text{GA}=1$ (see text for details).}
  \label{fig:lumfree}
\end{figure*}

We finish by exploring the implications that assuming at face value
the sterile solution of the Gallium anomaly would have on the
mechanism for energy production in the Sun as inferred from neutrino
data.  As mentioned above, \Nuc{pp} neutrinos yield the largest
contribution both to $L_\odot(\text{$\nu$-inferred})$ ($\gtrsim 90\%$)
and to the event rate of the Gallium solar experiments ($\gtrsim
55\%$), which poses the question of what should be the deviation from
the relation in Eq.~\eqref{eq:lumsum2} required for solar observations
to be compatible with the Gallium source experimental results.  In
order to quantitatively answer this question we have performed a SSM
independent analysis similar to the one described above, but without
imposing the prior in Eq.~\eqref{eq:priorLC}.  This allows us to
determine the level of compatibility between solar+KamLAND data and
Gallium source experiments as a function of the $\nu$-inferred solar
luminosity, by comparing the $\sin^2\theta_{14}$ range preferred by
each data set.  The results are shown in Fig.~\ref{fig:lumfree}.
In this analysis we have kept $f_\text{Ga}=1$ because of the strong
correlation (induced by $\Phi_{\Nuc{pp}}$) between the solar
luminosity and the event rate in the Gallium experiments, which
results in an almost complete degeneracy if both
$L_\odot(\text{$\nu$-inferred})$ and $f_\text{Ga}$ are left free to
vary at the same time.
In the left panel we plot the $1\sigma$, $2\sigma$, $3\sigma$ (1~dof)
ranges for $\theta_{14}$ (defined with respect to the global minimum)
allowed by the combined fit of solar and KamLAND data (for both
variants of the KamLAND analysis) as a function of the $\nu$-inferred
solar luminosity in units of the directly observed $L_\odot$.  For
comparison we show as horizontal grey bands the corresponding required
ranges to explain the Gallium anomaly.  As seen in the figure, for
either analysis, compatibility requires a substantial deviation of the
luminosity inferred from the solar neutrino observations with respect
to its value as directly determined.  This is further quantified in
the right panel, which shows the increase in $\chi^2$ when combining
together solar+KamLAND and Ga-source data (as a function of the
aforesaid luminosity ratio) with respect to the sum of the two
separate best-fits (so that by construction the minimum of each curve
yields the $\chi^2_\text{PG}$ value defined in Eq.~\eqref{eq:PGtest}).
As seen in the figure, for the KamLAND-RFC case the level of
compatibility is always higher than $2.8\sigma$, while in the
KamLAND-RFF case the compatibility only drops below the $2\sigma$
level if one allows $L_\odot(\text{$\nu$-inferred})$ to deviate by
more than $10\%$ from the directly determined solar luminosity.  In
other words, to accommodate the sterile neutrino interpretation of the
Gallium anomaly within the present observation of solar and KamLAND
neutrinos at better than the $2\sigma$ level, it is necessary to ($a$)
make no assumption on the normalization of the reactor antineutrino
fluxes, and ($b$) accept that more than 10\% of the energy produced in
the nuclear reactions in the Sun does \emph{not} result into observed
radiation --- despite the fact that, as mentioned before, the solar
radiated luminosity is directly determined with 0.34\% precision.

\section{Summary}
\label{sec:conclu}

In this work we have presented a variety of global analyses of solar
neutrino data (both alone and in combination with the KamLAND reactor
antineutrino results) in the framework of the 3+1 neutrino mixing
scenario commonly invoked to explain the results from Gallium source
experiments.  With these fits at hand, we have performed consistency
tests to assess the level of (in)compatibility between the range of
the sterile neutrino mixing preferred by each data set, and we have
studied the dependence of the results on the different assumptions
entering in the analysis.
All the fits considered here have shown compatibility (as measured by
the parameter goodness-of-fit) only at the $3\sigma$ level or higher.
This conclusion holds for analyses assuming solar neutrino fluxes as
predicted by any of the last generation Standard Solar Models, and
irrespective of the assumptions on the KamLAND reactor flux
normalization and on the gallium capture rate.  Relaxing the SSM
constraints on the solar fluxes ---~while still enforcing the relation
between the observed value of the solar radiated luminosity and the
total amount of thermal energy generated by the various
neutrino-emitting nuclear reactions~--- only improves the
compatibility to levels below $3\sigma$ when no prior knowledge in
either the gallium capture rate or the normalization of reactor
neutrino fluxes is assumed.  If the luminosity constraint is also
dropped, then it is formally possible to achieve a compatibility level
below $2\sigma$ as long as the flux of reactor antineutrinos is left
free, but such solution unavoidably requires that more than 10\% of
the energy produced together with neutrinos in the nuclear reactions
of the Sun does not result into observable radiation.

\section*{Acknowledgements}

This project is funded by USA-NSF grant PHY-1915093 and by the
European Union through the Horizon 2020 research and innovation
program (Marie Sk{\l}odowska-Curie grant agreement 860881-HIDDeN) and
the Horizon Europe programme (Marie Sk{\l}odowska-Curie Staff Exchange
grant agreement 101086085-ASYMMETRY).  It also receives support from
grants PID2020-\allowbreak 113644GB-\allowbreak I00,
PID2022-\allowbreak 142545NB-\allowbreak C21, ``Unit of Excellence
Maria de Maeztu 2020-2023'' award to the ICC-UB CEX2019-000918-M,
``Unit of Excellence Maria de Maeztu 2021-2025'' award to ICE
CEX2020-001058-M, grant IFT ``Centro de Excelencia Severo Ochoa''
CEX2020-001007-S funded by MCIN/AEI/\allowbreak 10.13039/\allowbreak
501100011033, as well as from grants 2021-SGR-249 and 2021-SGR-1526
(Generalitat de Catalunya).

\bibliographystyle{elsarticle-num}
\bibliography{references}

\end{document}